\documentclass[a4paper,UKenglish,cleveref,autoref,thm-restate]{lipics-v2021}

\usepackage{microtype}
\usepackage{graphicx}
\usepackage{booktabs}
\usepackage{multirow}
\usepackage{makecell}
\usepackage{array}
\usepackage{tabularx}
\usepackage{comment}
\usepackage{framed}
\usepackage{tcolorbox}
\tcbuselibrary{skins,breakable}
\usepackage{xspace}

\graphicspath{{figures/}}

\nolinenumbers

\newcommand*{\tool}{\texttt{TerraRepair}\@\xspace}

\title{\tool: Tool-Grounded LLM Agent for Repairing Infrastructure-as-Code}

\title{\tool: A Tool-Grounded LLM Agent for Infrastructure-as-Code Repair}

\authorrunning{Minase Mekete Mengistu, Juri Di Rocco, Phuong T. Nguyen, Davide Di Ruscio}

\author{Minase Mekete Mengistu}{University of L'Aquila, Italy}{minasemekete.mengistu@graduate.univaq.it}{https://orcid.org/0009-0006-9409-7529}{}

\author{Juri Di Rocco}{University of L'Aquila, Italy}{juri.dirocco@univaq.it}{https://orcid.org/0000-0002-7909-3902}{}

\author{Phuong T. Nguyen}{University of L'Aquila, Italy}{phuong.nguyen@univaq.it}{https://orcid.org/0000-0002-3666-4162}{}

\author{Davide Di Ruscio}{University of L'Aquila, Italy}{davide.diruscio@univaq.it}{https://orcid.org/0000-0002-5077-6793}{}

\ccsdesc[500]{Software and its engineering~Software maintenance tools}
\ccsdesc[300]{Software and its engineering~Software defect analysis}
\ccsdesc[300]{Security and privacy~Software security engineering}
\keywords{Infrastructure as Code, automated repair, large language models, cloud security}
\category{}
\relatedversion{}

\newtcolorbox{shadedbox}{
	drop shadow southeast,
	breakable,
	enhanced jigsaw,
	colback=white,
	boxrule=0.80pt,
	left=0.3em,
	right=0.3em,
	top=0.1em,
	bottom=0.05em
}

\EventEditors{Anonymous}
\EventNoEds{1}
\EventLongTitle{International Symposium on Empirical Software Engineering and Measurement (ESEM 2026)}
\EventShortTitle{ESEM 2026}
\EventAcronym{ESEM}
\EventYear{2026}
\EventDate{TBD}
\EventLocation{TBD}
\EventLogo{}
\SeriesVolume{XX}
\ArticleNo{XX}

\begin{document}
	
	\maketitle
\begin{abstract}
 \textbf{Background:} Infrastructure-as-Code (IaC) scanners detect cloud misconfigurations in Terraform and other IaC languages before deployment, but repairing the flagged configurations remains largely manual. Recent Large Language Model (LLM)-based repair approaches can repair some findings, but may hallucinate unsupported constructs or suppress warnings without fixing the issue. 

 \noindent\textbf{Aims:} 
 We study  whether tool grounding can improve LLM-based Terraform repair, and when a finding should be escalated because the required deployment-specific context is not available. 
 
 \noindent\textbf{Method:} We present \tool, a 
 prototype of a tool-grounded LLM agent for Terraform repair with structured escalation. \tool retrieves dependency context from Terraform references, consults the installed provider schema, and re-runs the scanner before returning a candidate repair. When the required context is absent, \tool escalates instead of fabricating a plausible fix. 
 
 \noindent\textbf{Results:} We evaluate our tool on two vulnerable-by-design Terraform repositories using two IaC security scanners, Checkov and Trivy, across AWS, Azure, and GCP. On the combined AWS benchmark, \tool improves scanner-verified fix rates from 26.6\% to 78.4\% on Checkov and from 44.8\% to 72.4\% on Trivy, compared with a controlled one-shot baseline. It also reduces the baseline's 44.8--73.6 percentage point (pp) claimed-vs-verified repair gap to under 5\,pp. In a sampled semantic audit, 78.9\% of \tool's scanner-verified AWS repairs are labeled as correct under a majority-vote protocol. 
 
 \noindent\textbf{Conclusions:} These emerging results show that tool grounding can substantially improve scanner-verified LLM-based IaC repair on the studied benchmarks, while missing deployment-specific context remains the main knowledge boundary for full autonomy. 

\end{abstract}

\section{Introduction} 
Infrastructure as Code (IaC) is widely used to provision and manage cloud infrastructure through declarative configuration files, such as Terraform modules written in HashiCorp Configuration Language (HCL)~\cite{nsa2024iac,hashicorp2023terraform}. This practice improves repeatability, reviewability, and automation, but it also moves cloud security and reliability risks into code~\cite{chiari2022static}. Misconfigured IaC can expose storage buckets, grant overly broad Identity and Access Management (IAM) permissions, disable encryption, or leave network access unrestricted. Prior work has studied such recurring weaknesses as IaC security smells~\cite{rahman2019seven}. 

IaC security scanners detect many of these issues before deployment by reporting policy violations over Terraform resources~\cite{chiari2022static,checkov,trivy}. However, detection is only the first step. Developers must still determine how to modify the affected resource, whether related resources must also be changed, and whether the repair is valid for the installed Terraform provider version. Many repairs also require deployment-specific context, such as Key Management Service (KMS) keys, Classless Inter-Domain Routing (CIDR) ranges, logging destinations, certificates, or least-privilege IAM scopes. 

LLMs offer a possible way to reduce this manual repair effort, following recent work on LLM-based program and vulnerability repair~\cite{pearce2023examining,jin2023inferfix}. In the IaC setting, Low et al.~\cite{low2024repairing} showed that LLMs can repair many Terraform scanner findings, especially when human-provided context is available. However, it has been shown that LLM-based IaC repair can hallucinate Terraform constructs, introduce validation errors, or clear scanner warnings without fixing the underlying security issue~\cite{low2024repairing}. This suggests that the main challenge is not only generating a plausible edit, but grounding the repair in the codebase, provider schema, scanner feedback, and available deployment context.

IaC repair differs from conventional program repair because IaC programs specify desired infrastructure state rather than local program behavior~\cite{saavedra2025intrafix}. A syntactically valid HCL edit may still be invalid for the installed provider schema, undeployable in the target cloud environment, or unsafe for the intended deployment. Prior work on semantic checks for cloud IaC shows that IaC programs that pass compilation can still fail at deployment time because syntax alone does not capture deployment-relevant requirements~\cite{qiu2024zodiac}. For this reason, scanner success is an incomplete repair oracle. An autonomous IaC repair system needs a way to decide when a repair is sufficiently grounded and when the finding should be escalated.

To bridge such a gap, we present \tool, a tool-grounded LLM agent for bounded Terraform repair. Given a scanner finding and the affected Terraform block, \tool treats repair as a bounded tool-using process. The agent can query a dependency graph for cross-resource context, inspect the installed Terraform provider schema, and re-run the scanner on candidate repairs. If the required information is absent from the codebase, 
\tool emits a structured escalation instead of fabricating a plausible repair. 

We evaluate \tool on two vulnerable-by-design Terraform repositories used in prior IaC repair research, TerraGoat~\cite{terragoat} and KaiMonkey~\cite{kaimonkey}. The evaluation uses Checkov and Trivy across AWS, Azure, and GCP configurations. We also construct a controlled one-shot baseline inspired by Low et al.'s autonomous first-pass prompt~\cite{low2024repairing}. This baseline is not a reproduction of their 2024 environment. It isolates direct prompting under the same model, scanner versions, finding population, patching logic, and scoring procedure used for \tool. On the combined AWS benchmark, \tool improves scanner-verified fix rates, outperforming the considered baseline for both Checkov and Trivy. 
The ablation and escalation analyses show that provider-schema grounding and dependency retrieval are the main contributors to repair effectiveness, while missing deployment-specific context remains the main knowledge boundary for full autonomy. 
 
As an emerging-results study, this paper makes the following contributions:
  \begin{itemize} 
  	\item We formulate Terraform security repair as a bounded tool-grounded process that returns either a scanner-verified candidate repair or a structured escalation. 
	\item The introduction of the claimed-vs-verified repair gap as an operational measure of unsupported repair claims, characterizing a key failure mode of direct LLM repair. 
  	\item The \tool tool as an empirical prototype that combines dependency-graph retrieval, provider-schema lookup, in-loop scanner verification, and escalation. To the best of our knowledge, this is the first tool of this type to tackle the issue. 
  	\item An initial empirical evaluation on vulnerable-by-design Terraform repositories across AWS, Azure, and GCP, comparing \tool with a controlled one-shot LLM baseline.  
	\item A replication package with the developed code and curated data has been published to foster open science~\cite{anonymous_replication_package}. 	
  \end{itemize}

\section{Related Work} 

IaC security scanners detect policy violations before deployment, but most prior IaC analysis work focuses on detection and verification rather than repair. Rahman et al.~\cite{rahman2019seven} identify recurring IaC security smells, including hard-coded secrets, empty passwords, and unrestricted network access. Chiari et al.~\cite{chiari2022static} survey static analysis for IaC, covering smell detection, defect prediction, and verification. InfraFix~\cite{saavedra2025intrafix} addresses IaC repair by inferring a desired system state, repairing a normalized intermediate representation, and propagating changes back to the original script. \tool instead repairs Terraform security findings reported by IaC scanners, using installed provider schemas, dependency references, scanner feedback, and escalation when required deployment context is absent. 

\tool also builds on automated program repair (APR) and recent LLM-based repair. APR generates patches with respect to evidence such as a test suite, crash, compiler error, or static-analysis warning~\cite{legoues2021automatic}, but a patch may overfit the repair oracle while failing to preserve intended behavior~\cite{smith2015cure}. This issue is especially relevant for security repair because clearing a warning does not imply that the underlying vulnerability has been fixed. Recent LLM repair work has explored zero-shot vulnerability repair, static-analysis-guided and retrieval-augmented repair, tool-using repair agents, and repository-history-aware repair~\cite{pearce2023examining,jin2023inferfix,bouzenia2025repairagent,shi2025hafixagent}. \tool follows this broader direction, but the artifact and oracle are different. It repairs Terraform configurations against scanner findings, provider schemas, dependency context, and explicit escalation conditions. 

Several studies 
apply LLMs to IaC and configuration-security tasks. TerraFormer~\cite{jana2026terraformer} studies Terraform generation and mutation from natural-language prompts using verifier-guided training over syntax, deployability, and policy compliance. LLMSecConfig~\cite{ye2025llmsecconfig} repairs Kubernetes container misconfigurations using static analysis, retrieval, prompting, and validation. Reyes et al.~\cite{reyes2025llm} construct a Terraform remediation dataset from Trivy findings and evaluate fine-tuning and in-context learning for open-source LLMs. Toprani and Madisetti~\cite{toprani2025agentic} propose a multi-agent RAG workflow for detecting CloudFormation vulnerabilities and producing remediation guidance. NSync~\cite{yang2025nsync} addresses Terraform reconciliation rather than security repair by using cloud API traces to infer out-of-band infrastructure changes and synthesize Terraform updates. Together, these works show that LLMs can support IaC generation, configuration repair, vulnerability remediation, and cloud-state reconciliation. 

Low et al.~\cite{low2024repairing} are the closest prior work. They repair Terraform scanner findings by passing the flagged block and scanner output to an LLM, with a human-provided second pass for remaining findings. Their results show that LLMs can remove many scanner warnings, but may also hallucinate Terraform constructs, introduce validation errors, or clear warnings without fixing the underlying issue. \tool targets this failure mode through provider-schema lookup, dependency retrieval, scanner feedback, and structured escalation.

\section{\tool: Proposed Approach}
\tool repairs Terraform scanner findings as a bounded tool-using process. Given a scanner finding, the flagged resource block, and the corresponding cause lines, \tool returns either a scanner-verified candidate repair or a structured escalation. The escalation is used when the required information is not available from the codebase, provider schema, or scanner feedback, or when the bounded repair process does not converge.
	
	\begin{figure}[t]
		\centering
		\includegraphics[width=1\linewidth]{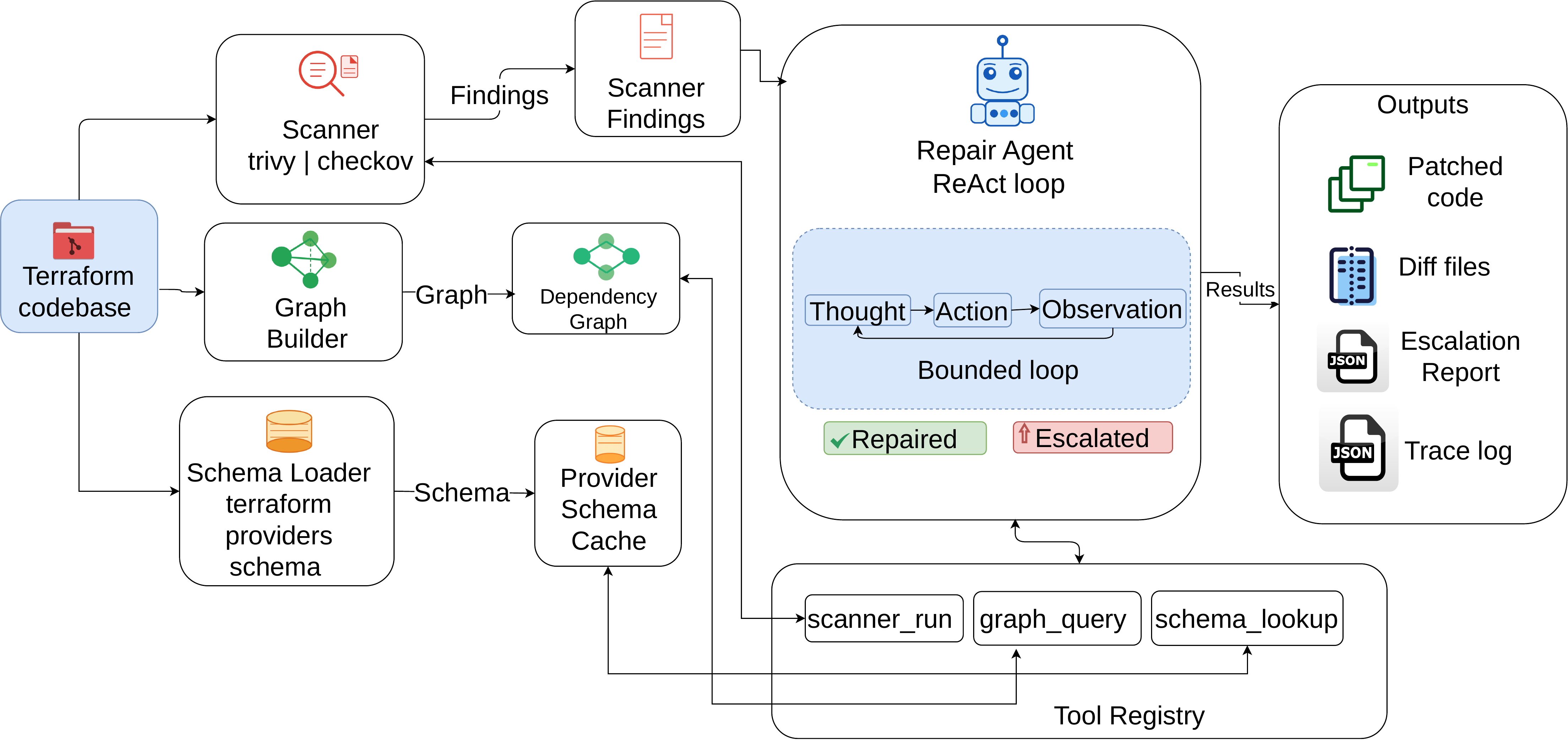}
		\caption{\tool architecture and repair workflow.}
		\label{fig:architecture}
	\end{figure}

The proposed approach is shown in Fig.~\ref{fig:architecture}. \tool first runs a security scanner to obtain misconfiguration findings. It  builds a dependency graph from HCL references in the codebase and loads the relevant Terraform provider schemas using the Terraform CLI provider-schema interface~\cite{terraformprovidersschema}. Each finding is then given to a single ReAct-style repair agent~\cite{yao2023react}. The agent receives the rule identifier, severity (if available), resolution hint, resource identifier, cause lines, and the complete original HCL block. We bound each repair attempt at ten agent steps. This fixed budget gives the agent room to query dependency context, inspect the provider schema, propose a patch, run the scanner, and revise when needed, while keeping repair attempts finite and comparable across findings. The agent has access to three tools. \textsc{graph\_query} retrieves cross-resource information from the dependency graph, e.g., KMS key ARNs, network identifiers, 
and security group references. \textsc{schema\_lookup} retrieves the installed provider schema for the target resource type, so that the agent can avoid unsupported attributes, invalid nested blocks, and assignments to provider-computed fields. \textsc{scanner\_run} re-runs the scanner on
the candidate repair and returns the result as feedback. If the scanner
still reports the original finding, the agent may revise the patch. If
the scanner no longer reports the original finding, the agent can
terminate and emit the candidate repair.

The repair prompt defines the agent role, available tools, stopping
criteria, and output schema. It enforces three constraints. First, the agent must consult the provider schema before using unfamiliar attributes or nested blocks. Second, it must obtain scanner verification before returning a candidate repair. Third, it must output exactly one repaired resource block. If a correct repair requires deployment-specific information that is absent from the codebase or provider schema, the agent is instructed to emit a structured escalation instead of fabricating a plausible repair. The full prompt templates, tool interfaces, and output schemas are included in the replication package~\cite{anonymous_replication_package}.

\section{Evaluation Methodology} \label{sec:methodology} 
The evaluation is organized around four research questions.

\noindent$\triangleright$ \textbf{\textit{RQ$_1$:}} How does \tool compare with direct one-shot LLM prompting?

\noindent$\triangleright$ \textbf{\textit{RQ$_2$:}} How does \tool vary across cloud providers?

\noindent$\triangleright$ \textbf{\textit{RQ$_3$:}} What is the individual contribution of each architectural component?

\noindent$\triangleright$ \textbf{\textit{RQ$_4$:}} What kinds of findings cannot be repaired autonomously?

We use two vulnerable-by-design Terraform repositories used in prior IaC repair research, namely TerraGoat (commit~\texttt{729f8da})~\cite{terragoat} and KaiMonkey (commit~\texttt{3feb0bf})~\cite{kaimonkey}. TerraGoat contains AWS, Azure, and GCP configurations. KaiMonkey contains four AWS slices covering compute, network, storage, and a cross-module Server-Side Request Forgery (SSRF) scenario. These seven benchmark slices are evaluated with Checkov and Trivy, yielding 14 dataset--scanner configurations. The primary comparison dataset is the combined AWS subset, which contains 227 Checkov findings and 168 Trivy findings, as shown in Table~\ref{tab:datasets}. We use AWS as the primary comparison because Low et al.'s Terraform repair evaluation used AWS. Azure and GCP assess cross-provider behavior.

\begin{table}[t]
	\caption{Benchmark datasets and baseline finding counts.}
	\label{tab:datasets}
	\centering\small
	\begin{tabular}{llrr}
		\toprule
		\textbf{Dataset} & \textbf{Provider}
		& \textbf{Checkov} & \textbf{Trivy}\\
		\midrule
		TerraGoat & AWS   & 158 & 113 \\
		TerraGoat & Azure & 124 &  90 \\
		TerraGoat & GCP   &  55 &  49 \\
		KaiMonkey & AWS   &  69 &  55 \\
		\midrule
		Combined AWS (primary) & --- & 227 & 168 \\
		Grand total            & --- & 406 & 307 \\
		\bottomrule
	\end{tabular}
\end{table}

We use Checkov~3.2.510, Trivy~0.69.3, \texttt{gpt-4o-mini-2024-07-18} at temperature~0.0~\cite{openai_temperature}, Terraform~CLI~0.14.11, and python-hcl2~7.3.1~\cite{pythonhcl2}. API inference may vary at temperature~0.0, so we repeat each \tool configuration three times. The AWS configurations resolve the AWS provider at versions 4.67.0 for TerraGoat and 3.76.1 for KaiMonkey. The non-AWS configurations resolve the AzureRM provider at version 4.68.0 and the Google provider at version 7.27.0. Low et al.~\cite{low2024repairing} used earlier scanner versions and also included Terrascan, which is now archived. Therefore, their published numbers are used only as historical context.

We construct a within-study
baseline based on Low et al.'s autonomous first-pass prompt~\cite{low2024repairing}.
The baseline uses the prompt from their Figure~2a verbatim and performs
one LLM call per finding, with no dependency-graph retrieval,
provider-schema lookup, or in-loop scanner verification. Both systems use
the same model, scanner versions, finding population, file-patching
logic, and post-run rescanning procedure. A parity check confirms that
both pipelines receive bit-identical scanner findings before repair.

The primary metric is the \emph{scanner-verified fix rate}, the
proportion of originally flagged findings that are no longer flagged when
the same scanner is re-run on the repaired output. We also report the
\emph{claimed-vs-verified repair gap}, defined as the claimed repair rate
minus the scanner-verified fix rate. For the baseline, a claimed repair
is an LLM response with \texttt{is\_fixed=true}. For \tool, it is a
final repaired status after in-loop \textsc{scanner\_run} verification.
Both systems are evaluated with the same final full-codebase rescan.

All \tool configurations are executed three times. The controlled
baseline is executed three times on the AWS configurations used for RQ1.
Each run first pools verified repairs across the relevant datasets, and
the reported value is the mean and standard deviation of these pooled
per-run rates.

We assess Terraform validity using differential
\texttt{terraform validate}, counting only newly introduced validation
errors. We assess semantic correctness using a sampled jury-of-judges
audit, informed by LLM-as-a-Judge work in software
engineering~\cite{he2026llmjudge}. From Run~3 AWS, whose fix rates are
closest to the three-run mean, we sample 171 of 303 scanner-verified
repairs using Cochran's formula with finite population correction~\cite{cochran1977}.
The sample is allocated proportionally across dataset--scanner strata,
and Wilson score intervals are used for the 95\% confidence
interval~\cite{wilson1927}.

Each sampled repair is assessed by the author and two LLM judges, Claude
Sonnet~4.5 (\texttt{claude-sonnet-4-5-20250929}) and GPT-5.4
(\texttt{gpt-5.4-2026-03-05}). The author inspects the original and
repaired code with repository context, while the LLM judges use a
conservative judging prompt that defaults to WRONG when the repair is
ambiguous. Final labels are determined by majority vote. Since two raters
are LLM judges and the human rater is an author, we treat this as a
semantic audit rather than ground-truth correctness.

For RQ3, we conduct a leave-one-out ablation on TerraGoat~AWS with
Checkov ($n=158$), comparing the full system with no
\textsc{graph\_query}, no \textsc{schema\_lookup}, and no
\textsc{scanner\_run}. Each configuration is executed three times. For
RQ4, we classify structured escalations into missing external context,
max-step termination, missing schema support, and unresolved Terraform
variable or local references.

\section{Experimental Results}
\label{sec:results}

\subsection{RQ$_1$: How does \tool compare with direct one-shot LLM prompting?}

The controlled one-shot baseline described in Section~\ref{sec:methodology} reproduces Low et al.'s autonomous first-pass prompt inside our repair harness. Table~\ref{tab:fixrate} reports the results on the combined AWS benchmark. The controlled baseline achieves a scanner-verified fix rate of \textbf{26.6\%\,$\pm$\,1.4\,pp} on Checkov and \textbf{44.8\%\,$\pm$\,1.4\,pp} on Trivy. \tool achieves \textbf{78.4\%\,$\pm$\,0.8\,pp} and \textbf{72.4\%\,$\pm$\,4.0\,pp}, respectively. This corresponds to an improvement of \textbf{+51.8\,pp} on Checkov and \textbf{+27.6\,pp} on Trivy over the one-shot baseline.

\begin{table}[t]
\caption{Scanner-verified fix rates on the combined AWS benchmark.
	Mean$\pm$std over three runs. Low et al.\ rows are historical context.}
	\label{tab:fixrate}
	\centering\small
	\setlength{\tabcolsep}{3pt}
	\begin{tabular}{llcc}
		\toprule
		\textbf{System} & \textbf{Approach}
		& \textbf{Checkov} & \textbf{Trivy}\\
		\midrule
		Low et al.\ GPT-3.5 pass~1 & autonomous, 2024 scanners
		& 9.0\% & 17.1\% \\
		Low et al.\ GPT-4 pass~1 & autonomous, 2024 scanners
		& 27.4\% & 44.2\% \\
		Low et al.\ GPT-4 pass~2 & human-assisted, 2024 scanners
		& 87.4\% & 67.3\% \\
		\midrule
		Controlled baseline & one-shot, \texttt{gpt-4o-mini}
		& 26.6\%\,$\pm$\,1.4\,pp
		& 44.8\%\,$\pm$\,1.4\,pp \\
		\textbf{\tool} & agent+tools, \texttt{gpt-4o-mini}
		& \textbf{78.4\%\,$\pm$\,0.8\,pp}
		& \textbf{72.4\%\,$\pm$\,4.0\,pp} \\
		\midrule
		\textbf{Gain over controlled baseline} & ---
		& \textbf{+51.8\,pp}
		& \textbf{+27.6\,pp} \\
		\bottomrule
	\end{tabular}
\end{table}

\begin{figure}[t]
	\centering
	\includegraphics[width=0.58\linewidth]{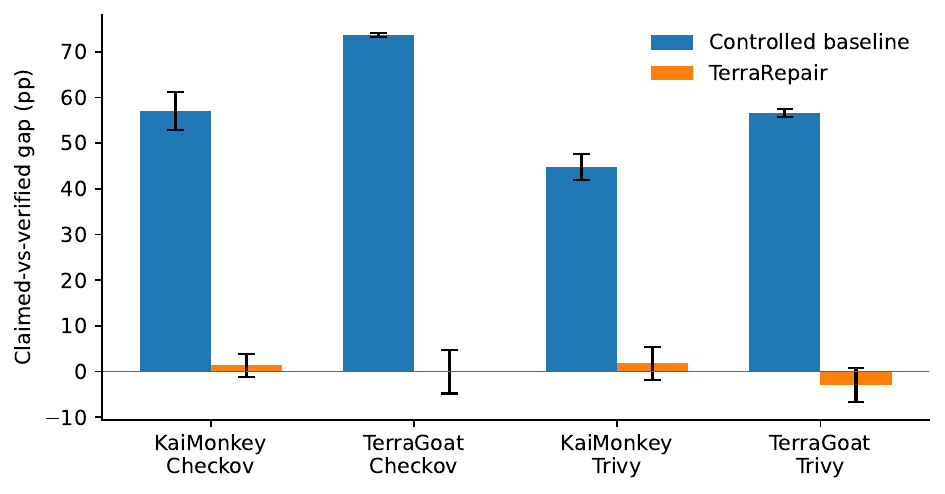}
	\caption{Claimed-vs-verified repair gap on AWS configurations. Values closer to zero indicate better calibration. Negative values indicate finding-level underclaiming.}
	\label{fig:claim-gap}
\end{figure}

The 
baseline also exposes a large gap between claimed and scanner-verified repairs. For the baseline, a claimed repair is an LLM response with \texttt{is\_fixed=true}. For \tool, a claimed repair is a final repaired status returned after in-loop \textsc{scanner\_run} verification. As shown in Fig.~\ref{fig:claim-gap}, the one-shot baseline has a claimed-vs-verified gap of 44.8--73.6\,pp across the AWS configurations. This means that the baseline frequently reports a repair as successful even though the same scanner still reports the original finding after patching and rescanning. \tool reduces this gap to between $-2.9$ and +1.8\,pp. A small negative gap means that the final scanner run cleared slightly more findings than \tool explicitly claimed to repair. This can happen because Terraform repairs are applied at the block level, while claims are recorded per scanner finding. If several findings refer to the same block, one repair may clear more than one finding. Thus, we interpret small negative gaps as finding-level underclaiming, not as evidence of additional semantic correctness. Positive gaps can occur when a repair passes \tool's in-loop scanner check but the finding reappears during the final full-codebase rescan. Scanner verification does not prove semantic correctness, but it reduces this specific unsupported-claim failure mode.

Repair validity is assessed at two additional levels. First, on the AWS benchmark and across all three runs, \tool introduces \emph{zero} new \texttt{terraform validate} errors. This provides evidence that the repaired configurations remain syntactically valid and
acceptable to Terraform validation on the primary benchmark, although
\texttt{terraform validate} is not a complete semantic oracle. Second,
in the sampled semantic audit, 135 of 171 scanner-verified AWS repairs
are judged semantically correct by majority vote, giving an estimate of
\textbf{78.9\%} (95\%~CI\,[72.2\%,\,84.4\%]). Inter-rater agreement is
moderate (Fleiss' $\kappa\!=\!0.54$), with 130/171 findings (76.0\%)
receiving unanimous verdicts. The judging prompt and rater-level labels are included in
the replication package~\cite{anonymous_replication_package}.

\vspace{.1cm}
\begin{shadedbox}
	\textbf{Answer to RQ$_1$.} \tool improves scanner-verified fix rate by +51.8\,pp on Checkov
and +27.6\,pp on Trivy, reduces the claimed-vs-verified gap to within $-2.9$ to +1.8\,pp, and has 78.9\% of sampled scanner-verified AWS repairs judged correct.
\end{shadedbox}

\subsection{RQ$_2$: How does \tool vary across cloud providers?}

Table~\ref{tab:multicloud} reports scanner-verified fix rates and validation errors across the four provider/dataset configurations. \tool achieves broadly similar scanner-verified fix rates across the evaluated settings, with 67.7\%--83.1\% on Checkov and 70.5\%--76.4\% on Trivy. There is no clear drop that can be explained by provider alone. The observed variation appears to be influenced by dataset composition, especially KaiMonkey's concentration of wildcard IAM policy repairs and TerraGoat~GCP's smaller finding population.

\begin{table}[t]
	\caption{Multi-cloud repair effectiveness and validity.
		Values are mean\,$\pm$\,std across three runs. Validation errors
		are combined totals across both scanners and three runs.}
	\label{tab:multicloud}
	\centering\small
	\setlength{\tabcolsep}{3pt}
	\begin{tabular}{llcccl}
		\toprule
		\textbf{Dataset} & \textbf{Prov.}
		& \textbf{Checkov} & \textbf{Trivy}
		& \textbf{Val.} & \textbf{Main validity issue}\\
		\midrule
		TerraGoat & AWS
		& 83.1\%\,$\pm$\,2.6
		& 70.5\%\,$\pm$\,2.8
		& 0 & none observed \\
		TerraGoat & Azure
		& 74.7\%\,$\pm$\,1.2
		& 72.2\%\,$\pm$\,2.2
		& 11 & missing required attrs. \\
		TerraGoat & GCP
		& 71.5\%\,$\pm$\,6.4
		& 71.4\%\,$\pm$\,2.0
		& 15 & invalid provider fields \\
		KaiMonkey & AWS
		& 67.7\%\,$\pm$\,6.7
		& 76.4\%\,$\pm$\,7.3
		& 0 & none observed \\
		\bottomrule
	\end{tabular}
\end{table}

Repair validity, however, shows a cross-provider limitation that scanner-verified fix rates alone hide. Across the three runs, all 26 newly introduced \texttt{terraform validate} errors occur in Azure and GCP repairs. These errors cluster around provider-specific schema failures. Missing required attributes and unsupported or removed attributes account for 19/26 errors (73.1\%). The remaining errors involve assignments to provider-computed fields, invalid enumeration values, and one block serialization failure. For example, for CKV\_AZURE\_7,\footnote{CKV\_AZURE\_7 checks that an Azure Kubernetes Service cluster has a network policy configured.} the agent generated \texttt{network\_profile \{ network\_policy = "calico" \}}, which satisfies Checkov's network-policy check but omits the required \texttt{network\_plugin} argument. Similarly, several GCP repairs used unsupported, removed, provider-computed, or invalid-enum attributes that were sufficient to address the scanner rule locally but invalid for the installed provider version. These cases show that provider-schema validity remains a separate repair requirement even when scanner-level behavior appears to generalize.


\vspace{.1cm}
\begin{shadedbox}
	\textbf{Answer to RQ$_2$.} \tool varies mainly in repair validity rather than scanner-verified fix rate. Scanner-verified fix rates remain broadly similar across AWS, Azure, and GCP, but all new validation errors occur in Azure and GCP repairs. Thus, scanner-level repair transfers better than schema-correct repair.
\end{shadedbox}

\subsection{RQ$_3$: What Is the Individual Contribution of Each Architectural Component?}

Table~\ref{tab:ablation} reports the leave-one-out ablation study on TerraGoat~AWS with Checkov. Here, \emph{Verified} denotes the post-run scanner-verified fix rate, and \emph{Agent} denotes the agent's in-loop repaired status. The largest measurable contributions come from \textsc{schema\_lookup} and \textsc{graph\_query}. Removing
\textsc{schema\_lookup} reduces the scanner-verified fix rate by 21.7\,pp and
more than doubles the mean number of escalations. This indicates that
training-time model knowledge is insufficient for reliable Terraform
repair. Removing \textsc{graph\_query} reduces the scanner-verified  fix rate by
15.6\,pp and similarly increases escalations. 

\begin{table}[t]
\caption{Leave-one-out ablation on TerraGoat AWS with Checkov (mean$\pm$std over three runs).}
	\label{tab:ablation}
	\centering\small
	\setlength{\tabcolsep}{3pt}
	\begin{tabular}{lcccc}
		\toprule
		\textbf{Configuration} & \textbf{Verified}
		& \textbf{$\Delta$} & \textbf{Agent}
		& \textbf{Escalations}\\
		\midrule
		Full system
		& 83.1\%\,$\pm$\,2.6 & ---
		& 83.1\%\,$\pm$\,2.2 & 26.7\,$\pm$\,3.5 \\
		No \textsc{graph\_query}
		& 67.5\%\,$\pm$\,4.5 & $-15.6$\,pp
		& 68.1\%\,$\pm$\,3.0 & 50.3\,$\pm$\,4.7 \\
		No \textsc{schema\_lookup}
		& 61.4\%\,$\pm$\,2.5 & $-21.7$\,pp
		& 59.1\%\,$\pm$\,2.4 & 64.7\,$\pm$\,3.8 \\
		No \textsc{scanner\_run}
		& 82.1\%\,$\pm$\,2.2 & $-1.1$\,pp
		& 81.4\%\,$\pm$\,2.9 & 29.3\,$\pm$\,4.5 \\
		\bottomrule
	\end{tabular}
\end{table}

Removing \textsc{scanner\_run} reduces the mean verified
fix rate by only 1.1\,pp. This small delta reflects the fact that
\textsc{scanner\_run} is primarily a verification and calibration tool
rather than a source of new repair context. It does not supply missing
resource identifiers or schema information, and therefore contributes
less to raw repair capability than \textsc{graph\_query} or
\textsc{schema\_lookup}. However, the controlled-baseline comparison in
RQ1 shows that scanner feedback is still important operationally as it
reduces unsupported repair claims by preventing repairs that still
trigger the original scanner finding from being reported as successful.


\vspace{.1cm}
\begin{shadedbox}
	\textbf{Answer to RQ$_3$.} \tool's scanner-verified fix rate appears to be driven primarily by live provider-schema grounding and graph-based dependency resolution, while scanner feedback mainly supports verified termination and calibrated success reporting.
\end{shadedbox}

\subsection{RQ$_4$: What Kinds of Findings Cannot Be Repaired Autonomously?} 
Across all 14 dataset--scanner configurations and three independent runs, \tool escalates a mean of 181.3 findings per run. This corresponds to a stable escalation rate of 25.4\% (std\,=\,1.3\,pp). Across the three runs, 544 findings are escalated in total. Table~\ref{tab:escalation} summarizes the escalation taxonomy. 

\begin{table}[t] 
	\caption{Escalation categories across all runs (mean escalation rate: 25.4\%).} 
	\label{tab:escalation} 
	\centering\small 
	\setlength{\tabcolsep}{3pt}
	 \begin{tabular}{lrrrrc}
	 	 \toprule 
	 	 \textbf{Category} & \textbf{R1} & \textbf{R2} & \textbf{R3} & \textbf{Mean} & \textbf{\%}\\ \midrule context\_ext. & 150 & 149 & 151 & 150.0 & 82.7\% \\ max\_step    &  31 &  21 &  16 &  22.7 & 12.5\% \\ schema\_unk. &   8 &   6 &   7 &   7.0 &  3.9\% \\ context\_var.&   3 &   1 &   1 &   1.7 &  0.9\% \\ \midrule \textbf{Total} & \textbf{192} & \textbf{177} & \textbf{175} & \textbf{181.3} & \textbf{100\%}\\ \bottomrule 
	\end{tabular}
 \end{table} 

The category labels denote missing external context (\texttt{context\_ext.}), max-step termination (\texttt{max\_step}), missing schema support (\texttt{schema\_unk.}), and unresolved Terraform variable or local references (\texttt{context\_var.}). The dominant category is missing external context (82.7\%). This includes absent cross-resource references and missing deployment-specific values, such as KMS ARNs, certificates, logging targets, secrets, disk-encryption sets, logging buckets, and network identifiers that cannot be inferred from the local codebase. Missing schema support and unresolved Terraform references account for another 4.8\% of escalations. 

A further 12.5\% of escalations are max-step terminations. These 
should not be interpreted as graceful agent escalations. The repair process reached the 10-step budget while the agent was still trying to converge. 
Heredoc-containing blocks such as IAM policies and \texttt{user\_data} scripts caused \textsc{scanner\_run} parse failures that the agent treated as repair syntax errors (20 cases). Scanner feedback did not identify the unsatisfied condition (20 cases). \textsc{schema\_lookup} lacked enough nested-block detail (14 cases). The LLM produced invalid tool-call formatting before any repair was attempted (7 cases). Finally, the agent exhausted steps on tool lookups or emitted \texttt{FINISH} as an invalid tool action (7 cases). These causes indicate repair-loop limitations rather than missing deployment intent.

\vspace{.1cm}
\begin{shadedbox}
	\textbf{Answer to RQ$_4$.} Most escalations are caused by unavailable or unresolved information, with missing external context alone accounting for 82.7\% of escalations. A smaller set of escalations comes from max-step termination, 
	where the bounded repair loop does not converge because of representation, scanner-feedback, or schema-detail limitations.
\end{shadedbox}

\section{Discussion}

\subsection{Limitations}

This study provides an initial empirical assessment of tool-grounded LLM repair for Terraform scanner findings and its results should be interpreted in light of various limitations. In particular, scanner-verified is not full semantic correctness, the semantic audit is not independent ground truth, the benchmarks are vulnerable-by-design repositories, and the results are tied to one repair model. 
Scanner-verified fix rate measures whether the original scanner finding
disappears after repair, but it is not a complete measure of repair
correctness. A scanner-verified repair may still be undeployable,
invalid for the intended cloud environment, or inconsistent with the
security policy. We therefore also report differential
\texttt{terraform validate} results and a sampled semantic audit. The
claimed-vs-verified repair gap is a reporting-calibration metric, not a
correctness metric.

\subsection{Threats to Validity}

\noindent$\triangleright$ \textbf{\textit{Internal validity.}}
The semantic audit uses one author and two LLM judges, which may
introduce author bias and model-judging bias. We mitigate this threat by
using a structured judging prompt, reporting inter-rater agreement, and
using majority vote. However, the audit should still be interpreted as
an estimate under the jury protocol rather than as ground-truth repair
correctness.

\noindent$\triangleright$ \textbf{\textit{External validity.}}
The evaluation uses TerraGoat and KaiMonkey, which are public and
reproducible but vulnerable-by-design repositories. Production IaC may
include private modules, custom providers, remote state, provider
aliases, dynamic blocks, workspace-specific variables, and
organization-specific deployment conventions. The experiments also use
one repair model, \texttt{gpt-4o-mini-2024-07-18}, at temperature~0.0.
Therefore, the fix rates, claim gaps, and ablation ordering should not
be interpreted as model-independent properties of LLM-based IaC repair.

\noindent$\triangleright$ \textbf{\textit{Conclusion validity.}}
The primary RQ1 comparison is a controlled within-study comparison, but
it is an architecture-level comparison rather than an ablation of
individual mechanisms. \tool differs from the one-shot baseline in
tool grounding, chain-repair within blocks, in-loop scanner
verification, and agent-oriented prompting. The leave-one-out ablation
isolates the three tools within \tool, but does not separately
isolate prompting and repair-ordering differences. 

\section{Conclusion and Future Work} 
This paper presented \tool, a tool-grounded LLM agent for bounded Terraform repair. \tool retrieves dependency context, inspects the installed provider schema, and re-runs the scanner before returning a candidate repair. When required information is absent, it emits a structured escalation instead of fabricating a plausible fix. 

In a controlled comparison with a one-shot baseline using the same model, scanner versions, finding population, patching logic, and scoring procedure, \tool improves scanner-verified fix rate 
on Checkov and 
on Trivy on the combined AWS benchmark. It also reduces the baseline's 
repair gap. 
The ablation, audit, and escalation analyses suggest that provider-schema grounding and dependency retrieval are important contributors to repair effectiveness, while missing deployment-specific context appears to be a major knowledge boundary for autonomous repair. 

The long-term goal is to extend \tool to production-like Terraform repositories. Toward this goal, future work will address the observed failure modes by improving schema detail for nested HCL blocks and tool interfaces for heredoc-containing repairs. Future work will also integrate organization-specific knowledge sources, such as approved KMS keys, logging destinations, certificate inventories, and IAM policy templates, to support context-aware repair. Finally, semantic correctness should be assessed through independent review by IaC and cloud-security practitioners.

\section{Data Availability}
The replication package supporting this study is available in an anonymized public repository~\cite{anonymous_replication_package}. It includes the \tool source code, evaluation scripts, semantic-audit artifacts, escalation labels, and instructions for reproducing the results. The benchmark datasets are available from the TerraGoat and KaiMonkey repositories at the commits listed in Section~\ref{sec:methodology}.

\section{Acknowledgments}
This paper has been partially supported by the MOSAICO project (Management, Orchestration and Supervision of AI-agent COmmunities for reliable AI in software engineering) that has received funding from the European Union under the Horizon Research and Innovation Action (Grant Agreement No. 101189664). 

\bibliography{main}
	
\end{document}